\documentstyle[preprint,eqsecnum,aps]{revtex}

\begin{document}
\draft
\title{EXPLICIT RELATIVISTIC VORTEX SOLUTIONS\\
FOR COOL TWO-CONSTITUENT SUPERFLUID DYNAMICS}
\author{Brandon Carter and David Langlois}
\address{D\'epartement d'Astrophysique Relativiste
et de Cosmologie, C.N.R.S.,\\
Observatoire de Paris, 92195 Meudon, France.\\
Racah Institute of Physics, The Hebrew University,\\
Givat Ram, 91904, Israel.}
\date{\today}
\maketitle

\begin{abstract}
We give a class of explicit solutions for the stationary and cylindrically
symmetric vortex configurations for a ``cool'' two-component superfluid
(i.e. superfluid with an ideal gas of phonons). Each solution is
characterized only by a set of (true) constants of integration. We then compute
the effective asymptotic contribution of the vortex to
the stress energy tensor by comparison with a uniform reference state
without vortex.
\end{abstract}

\section{Introduction.}
\label{sec:1}

The subject of investigation in the present work is the class of
vortex type (stationary cylindrically symmetric) configurations
for the relativistic generalisation of Landau's two constituent
superfluid theory, using the recently derived Lagrangian formulation\cite{1}
in which the independent variables are the superfluid phase scalar
$\varphi$ and the entropy current vector $s^\rho$.  More particularly,
it will be shown that the complete set of such  vortex solutions is
obtainable in analytic form in case of the ``cool" phonon dominated
limit regime. The required form of the Lagrangian for this ``cool"
regime has been found\cite{2} to be simply given by
\begin{equation}
{\cal L}=P-3\psi \ ,\label{eq:1.1}
\end{equation}
where $P$ is the ``cold" pressure function, depending just on the superfluid
momentum covector
\begin{equation}
\mu_\rho=\hbar\nabla\!_\rho\varphi\ ,\label{eq:1.2}
\end{equation}
which governs the zero temperature limit for which the entropy current
vanishes, while the thermal contribution $\psi$ represents the generalised
pressure of the phonon gas, which is given by a simple algebraic
expression (4.3) involving $s^\rho$ as well as $\mu_\rho$ and the ``cold"
sound speed $c_{_{\rm S}}$ determined by the pressure function $P$.

It is to be remarked that whereas accurate treatment of non-stationary
configurations with non-zero temperature would require allowance for
viscosity of the entropy current, however,  no loss of accuracy
will be entailed by the use of the strictly conservative treatment
as in Landau's original model\cite{3} for treating the stationary
equilibrium configurations under consideration here.
Indeed under these circumstances there will
in any case be no dissipation, the only effect of viscosity being to ensure
that the ``normal" part of the flow is constrained to
have a configuration that is {\it rigid}.

Unlike the neglect of dissipation that is implicit in our use of a
strictly Lagrangian formulation, the limitation that the states under
consideration here be restricted to the ``cool" regime governed by a
Lagrangian of the particular form (\ref{eq:1.1}) in the sense explained above
represents a significant physical limitation. It is provisionally necessary
to postpone detailed quantitative analysis of states in the physically very
interesting ``warm" regime nearer to the phase transition (beyond which lies
the regime of ``hot" states in which superfluidity is absent altogether).
The reason is not merely the expectation that the equation of state
governing the explicit form of the dynamical equations in the ``warm" regime
would be too complicated to be easily tractible. A more compelling obstacle
is the consideration that an appropriate ``warm" generalisation
 of the ``cool" equation of state (\ref{eq:1.1}) is not
not yet available at all. Although extrapolation beyond the``cool" regime
is not yet possible for all the explicit quantitative results obtained below,
it will nevertheless be found possible to obtain many useful qualitative
results that remain valid even in the ``warm" regime, since as shown in
Section~\ref{sec:3}, the relevant conservation laws provide sufficiently many
first integrals to solve the differential part of the vortex problem
completely in the general case. One is thus left with a purely algebraic
problem that remains intractible in the general case, but that is
easily solved in the ``cool" limit.

The formalism that will be used in the present work is fully covariant in the
general, not just special, relativistic sense, but since the kinds of
(neutron star or laboratory Helium) vortex that are envisaged are of limited
scale it will be justifiable to ignore gravitation, i.e. to use a simple
Minkowski space background, in the actual application to explicit solutions.
As well as being obviously important for applications such as neutron star
interiors\cite{4} in which relativistic effects are actually quite large, the
use
of a covariant formulation of superfluidity theory is also advantageous as a
source of supplementary physical insight even for the familiar laboratory
example of liquid Helium-4 in which relativistic corrections are
quantitatively negligible\cite{5}.

\section{The equations of motion for two-constituent superfluid dynamics.}
\label{sec:2}

Quite generally (not just in the``cool" regime) the dynamical equations for
the generalised Landau theory are determined by the Lagrangian ${\cal L}$
in terms of
the entropy vector $s^\rho$ and the momentum covector $\mu_\rho$ given by
(\ref{eq:1.1}) together with their dynamical conjugates, namely the
 thermal 4-momentum covector $\Theta_\rho$ and the particle number
current vector $n^\rho$ that are constructed\cite{1} from
the Lagrangian according to the infinitesimal variation formula
\begin{equation}
d{\cal L}=\Theta_\rho ds^\rho-n^\rho d\mu_\rho \ . \label{eq:2.1}
\end{equation}
In terms of these quantities, the equations of motion consist just of
the thermal momentum evolution equation
\begin{equation}
s^\rho\nabla\!_{[\rho}\Theta_{\sigma]}=0 \label{eq:2.2}
\end{equation}
(using square brackets to denote index antisymmetrisation) together
with the usual particle and entropy conservation laws
\begin{equation}
\nabla\!_\rho n^\rho=0 \label{eq:2.3}
\end{equation}
and
\begin{equation}
\nabla\!_\rho s^\rho=0 \ .\label{eq:2.4}
\end{equation}

The formulation that has just been summarised has the technical advantage of
being particularly economical in so much as it involves only 5 independent
component variables, namely the phase scalar $\varphi$ and the 4 independent
components of the entropy current vector $s^\rho$. This feature of economy has
recently been exploited for the purpose of setting up a correspondingly
economical Hamiltonian formulation of the theory\cite{6}, and it has also been
exploited as a guide to the formulation of an analagously economical theory
for describing thermal effects in superconducting cosmic strings\cite{7}.

The dynamic equations (\ref{eq:1.2}), (\ref{eq:2.2}), (\ref{eq:2.3}),
(\ref{eq:2.4}) entail the corresponding
pseudo (in flat space strict) energy momentum conservation law
\begin{equation}
\nabla\!_\rho T^{\rho\sigma}=0\label{eq:2.5}
\end{equation}
where the stress momentum energy density tensor is given in terms of
the corresponding generalised pressure function
\begin{equation}
\Psi={\cal L}-\Theta_\rho s^\rho\  \label{eq:2.6}
\end{equation}
by
\begin{equation}
 T^\rho_{\ \sigma}=n^\rho\mu_\sigma+s^\rho\Theta_\sigma+\Psi g^\rho_{\
        \sigma}\ .  \label{eq:2.7}
\end{equation}

An obviously convenient way of choosing the 3 independent variables in a
fundamental state function of the form characterised by (\ref{eq:2.1})
will be to take them to consist of the thermal rest frame entropy
density $s$ as given by
\begin{equation}
c^2 s^2=-s_\rho s^\rho \ .\label{eq:2.8}
\end{equation}
the new cross product variable $y$ given by
\begin{equation}
c^2 y^2 =-\mu_\rho s^\rho \ ,\label{eq:2.9}
\end{equation}
together with the effective mass variable $\mu$ given by
\begin{equation}
c^2\mu^2=-\mu_\rho\mu^\rho \ .\label{eq:2.10}
\end{equation}

It can then be seen that the secondary variables $n^\rho$ and
$\Theta_\rho$ will be given in terms of the primary variables $\mu_\rho$
and $s^\rho$ of this formulation by
\begin{equation}
n^\rho=\Phi^2\big(\mu^\rho -{\cal A} s^\rho\big)\  ,\qquad\quad
\Theta_\rho=\Phi^2\big( {\cal K}s_\rho + {\cal A} \mu_\rho\big) \
  ,\label{eq:2.11}
\end{equation}
where the relevant dilation, determinant and anomally coefficients $\Phi^2$,
${\cal K}$, and ${\cal A}$ are given by the partial differentiation formulae
\begin{equation}
c^2\Phi^2={1\over\mu}{\partial {\cal L}\over\partial\mu}\ ,\qquad \quad
c^2\Phi^2{\cal K}=-{1\over s}{\partial {\cal L}\over\partial s}\ ,\qquad\quad
c^2\Phi^2{\cal A}=-{1\over 2 y}{\partial {\cal L}\over\partial y}\ .
\label{eq:2.12}
\end{equation}
In terms of these variables, (2.7) can be rewritten as
\begin{equation}
T^{\rho\sigma}=\Phi^2\big(\mu^\rho\mu^\sigma+{\cal K}s^\rho s^\sigma
\big)+\Psi g^{\rho\sigma}\ .\label{eq:2.13}
\end{equation}

\section{The class of strong equilibrium states of a cylindrical vortex.}
\label{sec:3}

The class of vortex configurations to be dealt with here is of the maximally
symmetric type, as characterised by both stationarity and cylindrical
symmetry. This means that there are three independent commuting Killing
vector symmetry generators, $k^\mu$, $\ell^\mu$, and $m^\mu$ say, which may
be taken to be the generators of time translations, longitudinal space
translations (parallel to the axis) and axial rotations, respectively
corresponding to ignoreable coordinates $t$, $z$, and $\phi$ say, where the
latter (like the superfluid phase variable $\varphi$) is periodic with
period $2\pi$. Within this stationary cylindrical category, the class of
solutions that will be dealt with here is characterised by the condition of
equilibrium in the {\it strong} sense, which  is to be understood here as
implying exclusion of the (experimentally possible, but under natural
conditions unlikely) presence of any source or sink of energy or fluid flux
in the central core of the vortex (where the superfluidity conditions
inevitably breakdown).

In view of the conservation laws (2.3) and (2.4), the
condition of equilibrium in this strong sense evidently entails that the
flow has no radial components and so is necessarily {\it helical }:
this means that the current vectors will be confined to the timelike
hypersurface generated by the Killing vectors, so that they will be
expressible in the form
\begin{eqnarray}
   n^\rho&=&\nu(k^\rho+v\ell^\rho+\omega m^\rho)  \ , \label{eq:3.1}\\
  s^\rho&=&\sigma(k^\rho+V\ell^\rho+\Omega m^\rho)\ . \label{eq:3.2}
\end{eqnarray}
The class of such helical flow configurations includes the specially simple
and important subclass of {\it circular} flow configurations, namely those
for which the longitudinal translation velocities $v$ and $V$ can be taken
to be zero (for a suitably adjusted choice of $k^\rho$ and $\ell^\rho$), as
is the case in the kinds of vortex that occurs most commonly in practice.
Whereas both the amplitude factors, $\nu$ and $\sigma$, and in the particle
current case also the translation velocity $v$ (if relevant) and the angular
velocity $\omega$, can be expected to be radially dependent (though they
must of course be independent of the ignoreable coordinates, $t,z,\phi$), it
is however implicit in the ``strong" equilibrium condition that in so far as
the entropy constituent is concerned the corresponding ``normal" translation
velocity $V$ (if relevant), and the corresponding ``normal" angular velocity
$\Omega$ have to be strictly {\it uniform}, i.e. constant in the radial
direction as well. This latter requirement expresses the condition that the
``normal" constituent flux should be {\it rigid} in the sense of having its
current vector aligned with a Killing vector field, $\bar k^\rho$ say,  whose
explicit form will be given by
\begin{equation}
\bar k^\rho=k^\rho+V\ell^\rho+\Omega m^\rho=\sigma^{-1}s^\rho\ .\label{eq:3.3}
\end{equation}
The reason why it is implicit in the ``strong" equilibrium condition that
the ``normal" constituent of the flow should be rigid, is that in a
realistic treatment ``normality" implies the presence of at least a small
amount of viscosity which for non-rigid motion would cause dissipation. This
could be compatible with stationarity only in the presence of a radial flux
of entropy and hence also of energy, which would need to be supplied by the
core even if the only heat sink were at the outer boundary.

As well as the simplifications involved in (\ref{eq:3.1}) and more
particularly in (\ref{eq:3.3}), it is to be remarked that the
restriction that we are only considering states of equilibrium in the
strong sense --~requiring a non dissipative flow configuration~-- has
the practical advantage that there is no loss of realism in our use of a
strictly conservative theory, of the kind summarised in Section 1:
there would be no point in including the ``normal" viscosity terms that
are relevant in other contexts, because in the states treated here their
presence would have no effect.

The condition that the ``normal" part of the flow is rigid makes it
subject to a relativistic generalisation\cite{8} of a (Jacobi type)
variant of the classical Bernoulli theorem. The condition that a Killing
vector, $\bar k^\rho$ say, generates a symmetry of flow means that for
any physical well defined field (though not of course for a gauge dependent
quantity such as the superfluid phase variable $\varphi$) the
corresponding Lie derivative should vanish.  In the case of the thermal
momentum covector $\Theta_\rho$ this symmetry condition is expressible
as
\begin{equation}
2\bar k^\rho\nabla\!_{[\rho}\Theta_{\sigma]} +\nabla\!_\sigma(\bar k^\rho
   \Theta_\rho) =0 \ .\label{eq:3.4}
\end{equation}
The condition that the flow satisfies a rigidity condition of the form
(\ref{eq:3.3}) can now be used to rewrite this Lie invariance condition
in terms of the relevant current vector in the more specialised form
\begin{equation}
2s^\rho\nabla\!_{[\rho}\Theta_{\sigma]} =
-\sigma\nabla\!_\sigma(\bar k^\rho\Theta_\rho) \ .\label{eq:3.5}
\end{equation}
Up to this point the reasonning has been purely kinematic. However if we
now use the fact that the momentum covector in question is subject to a
dynamical equation of the standard form (\ref{eq:2.2}) it can be seen that
the left hand side of (\ref{eq:3.5}) will simply vanish, leaving us with a
uniformity condition that provides us with a first integral of the motion
in the form
\begin{equation}
\bar k^\rho\Theta_\rho=-\bar\Theta  \ , \label{eq:3.6}
\end{equation}
where $\bar\Theta$ is a Jacobi-Bernouilli type constant, which is
interpretable as an {\it effective temperature} that (like the quantities
$V$ and $\Omega$ introduced above) is {\it uniform} throughout. The existence
of this uniform effective temperature (as measures with respect to the
rigidly corotating frame of the ``normal" flow) is interpretable as
the relevant application of the ``zeroth law of thermodynamics".

Although the particle current does not satisfy a rigidity condition, the
fact that the associated dynamical equation is not just of the standard form
but has the more restrictive form of the irrotationality condition
(\ref{eq:1.2})
provides us with not merely one but three more independent Bernouilli type
constants, corresponding respectively to the three independent Killing
vectors. One way to see this is to start by substituting $\mu_\rho$ in the
place of $\Theta_\rho$ in the Lie invariance condition (\ref{eq:3.4}).
Since it is obvious that the dynamic condition (\ref{eq:1.2}) will then
immediately anullate the first term, one is again left with a uniformity
condition that provides another Bernouilli type constant, $\bar E$,
which is interpretable as the {\it energy} per particle with respect to
the rigidly rotating ``normal" frame, and which is given by
\begin{equation}
\bar k^\rho\mu_\rho=-\bar E .\label{eq:3.7}
\end{equation}
Furthermore, since this latter result is not of the Jacobi type in that
its derivation did not actually depend on the rigid corotation property
that singles out the combination $\bar k^\rho$, it is clear that an
analogous, ordinary rather than Jacobi type, Bernouilli constant can be
obtained by replacing $\bar k^\rho$ with any other symmetry generating
Killing vector.  We thus obtain not merely one but three more
independent Bernouilli type constants, $E$, $L$, and $M$ say,
corresponding to the three independent Killing vectors, $k^\rho$,
$\ell^\rho$, and $m^\rho$, and interpretable respectively as
representing the {\it energy, longitudinal momentum}, and {\it angular
momentum}, per particle, which will be given by
\begin{equation}
k^\rho\mu_\rho=-E\ , \hskip 1 cm \ell^\rho\mu_\rho=L\ , \hskip 1 cm
m^\rho\mu_\rho=M \ .\label{eq:3.8}
\end{equation}
The prototype example (\ref{eq:3.7}) is of course not independent of
these, but is given in terms of them as the constant coefficient linear
combination
\begin{equation}
\bar E= E-VL-\Omega M \ . \label{eq:3.9}
\end{equation}

Up to this point everything that has been done is fully general relativistic
in the sense of being applicable with respect to an arbitrary system of
coordinates $\{x^\rho\}$  for an arbitrarily curved space time metric
$g_{\rho\sigma}$ . If we wished to allow for self-gravitation we would still
be left with a non-trivial differential system of Einstein type equations to
be solved. However if, as will be sufficient in all the most obvious
applications, we are willing to treat the gravitational spacetime background
as fixed in advance, independently of any feedback from the superfluid
system, then it is apparent that the work carried out already above has been
sufficient to fully integrate all the relevant differential equations. To
see this, it suffices to notice that, as the helicality property implies the
absence of any radial components, each of the pair of currents involved has
only three instead of four independent components, which means that the
system is characterised locally by a total of only 6 independent field
components. Therefore, in order to solve the corresponding differential
system, it suffices to obtain a corresponding set of just 6 first integral
constants. Precisely such a set is provided by the work that has just been
described: assembling them all together the complete sextet of constants can
be listed as the temperature $\bar\Theta$ and the linear and angular
velocities $V$ and $\Omega$ associated with the ``normal" part of the
system, together with the energy $E$ (or its corotating counterpart $\bar E$)
and the linear and angular momenta $L$ and $M$ associated with the
superfluid part.

It is to be remarked that as far as the physical characterisation of the
solution is concerned only 5 of these constants are independent, since by a
longitudinal Lorentz adjustment either $V$ or $L$ given any desired value
without loss of generality, the most mathematically convenient choice in the
present formulation being to use the ``superfluid frame" in which $L=0$.

It is also to be remarked that whereas the first five members of the sextet
that has just been listed can take values in a continuous range, the
quantisation condition on the phase variable $\varphi$ in (\ref{eq:1.2})
entails of course that the last one, $M$, should be restricted to
integral multiples of $\hbar$.  Moreover, since vortices with higher
values of the relevant winding number will typically be unstable, it is
in common circumstances sufficient to consider only the almost trivial
case with $M=0$ (which might be deemed unworthy of description as proper
vortex) and the first proper vortex possibility as given (subject to a
judicious choice of sign convention) by $M=\hbar$.

Although, in a given background, the differential part of the problem is in
principle solved by the foregoing derivation of the sextet of constants
$\bar\Theta$, $V$, $\Omega$, $E$, $L$, $M$, there remains what
--~depending on the complexity of the relevant equation of state~--
may be a highly non-trivial algebraic problem that still has to be solved if
one wishes to obtain all the other important non-uniformly distributed
physical field quantities involved in explicit form as variable functions of
a suitably defined cylindrical radial coordinate $r$ say, which can be taken
to be the only independent variable in each solution.

Since such superfluid vortex configurations as can in practice be set up
artificially under laboratory conditions, or as can be expected to  occur
naturally in neutron stars, are all characterised by dimensions that
are small compared with the length scales over which self gravitational
effects become important, they should be adequately describable in terms of
a local gravitational background that is not merely given in advance (which
is all that is needed for complete integrability as described above) but
that can be taken more particularly to be {\it flat}. This allows us to
work with a system  of the standard cylindrical form
$\{x^{_0},x^{_1},x^{_2},x^{_3}\} =\{t,z,\phi,r\}$,
such that only the radius coordinate $r$ is non-ignoreable, while the
others are the ignoreable coordinates $t$, $z$, $\phi$, mentionned above
in association with the Killing vectors $k^\rho$, $l^\rho$, $m^\mu$,
so that the metric will be given by the familiar formula
\begin{equation}
g_{\rho\sigma}dx^\rho dx^\sigma=-c^2 dt^2+dz^2+r^2d\phi^2+dr^2\ .
\label{eq:3.10}
\end{equation}
With respect to this coordinate system the independent Killing vector
symmetry generators will be given by
\begin{equation}
k^\mu\leftrightarrow (1,0,0,0)\ , \qquad \ell^\mu\leftrightarrow
(0,1,0,0)\ , \qquad m^\mu\leftrightarrow (0,0,1,0)\ .\label{eq:3.11}
\end{equation}
The corresponding components of the primary momentum covector and
current vector in terms of which the formulation based on the Lagrangian
${\cal L}$ is formulated will be given by
\begin{equation}
\mu_\rho\leftrightarrow (-E,L,M,0)\ , \qquad\quad
   s^\rho\leftrightarrow \sigma(1,V,\Omega,0)\ .\label{eq:3.12}
\end{equation}
It is thus apparent that, in order to obtain the complete solution for
all the relevant dynamical variables, all that is still needed is to
find the functional dependence of the single variable $\sigma$ on $r$.

It can immediately be seen that the required function $\sigma$ is directly
proportional to the quantity $y^2$ as defined by (\ref{eq:2.9}), which
can be evaluated using (\ref{eq:3.12}) simply as
\begin{equation}
       y^2={\bar E\over c^2}\sigma \ . \label{eq:3.13}
\end{equation}
It can similarly be seen that $\sigma$ is also related by a simple
though radially variable proportionality factor to the quantity $s$ as
defined by (\ref{eq:2.8}), which can be evaluated using (\ref{eq:3.12}) as
\begin{equation}
    s^2=\Big(1-{V^2\over c^2}-{\Omega^2 r^2\over c^2}\Big)\sigma^2\
 .\label{eq:3.14}
\end{equation}
It follows that although it is still necessary to find out how
$\sigma$ depends on $r$ in order to obtain the corresponding radial
dependence of the separate scalar state variables $s$ and $y$, the ratio
$y^2/s$ has a radial dependence that is immediately expressible in
terms of the Lorentz factor
\begin{equation}
\Gamma = \Big(1-{V^2\over c^2}-{\Omega^2 r^2\over c^2}\Big)^{-1/2} \ ,
 \label{eq:3.15}
\end{equation}
associated with the rigid velocity distribution of the ``normal" flow by
the formula
\begin{equation}
{y^2\over s}={\bar E{\Gamma}\over c^2}\ .\label{eq:3.16}
\end{equation}

Although (\ref{eq:3.12}) provides only the ratio (\ref{eq:3.16}), but not
the absolute values, of $s$ and $y$, it is more helpful for the third of
the scalar fields that is needed for evaluating the Lagrangian function,
namely the effective mass function $\mu$ as defined by (\ref{eq:2.10}),
which can be seen to be given explicitly as a function of $r$ by the
easily memorable formula
\begin{equation}
c^2\mu^2= {E^2\over c^2}-L^2-{M^2\over r^2}\ . \label{eq:3.17}
\end{equation}

It is evident that in the ``cold" limit for which the entropy current
vanishes the explicit solution is directly available in terms of just the 3
constants involved in (\ref{eq:3.17}) namely $E$, $M$ and $L$ (of which
the last is physically redundant, being adjustable to zero by a
longitudinal Lorentz transformation) since in that case the Lagrangian
depends only on the single scalar $\mu$.  It is noteworthy that while
the very simple formula (\ref{eq:3.17}) provides the complete solution
to the (strong) equilibrium problem for a relativistic vortex in the
cold limit, it remains formally valid without change, although it is no
longer sufficient to provide the complete solution all by itself, in the
generic ``warm superfluid" case.

To complete the solution of the vortex problem in the ``warm" case, we must
make use of the remaining (sixth) constant of integration, which (unlike the
other five) does not appear in (\ref{eq:3.12}), namely the corotating
``normal frame" temperature $\bar\Theta$ as introduced in
(\ref{eq:3.6}).  With the aid of (\ref{eq:2.11}) it can be seen that
(\ref{eq:3.6}) can be rewritten in the form
\begin{equation}
\bar\Theta\, c^2 y^2=\bar E\big( \Psi-{\cal L}\big)\ ,\label{eq:3.18}
\end{equation}
where the bracketted function on the right is given explicitly by
\begin{equation}
\Psi- {\cal L}=-s{\partial{\cal L}\over\partial s} -
{y\over 2}{\partial{\cal L}\over\partial y}\ .\label{eq:3.19}
\end{equation}

In principle, since $\mu$ is given directly by (\ref{eq:3.17}), all that
remains to be done to obtain the other two state variables $y$ and $s$
(and hence everything else) as functions of $r$ is to carry out the
simultaneous solution of the pair of algebraic equations (\ref{eq:3.16})
and (\ref{eq:3.18}) using the explicit radial dependence given by
(\ref{eq:3.15}) and (\ref{eq:3.17}).  In practice however, whereas the
solution is immediate in the cold limit (for which $y$ and $s$ vanish so
that (\ref{eq:3.17}) suffices by itself), in the generic ``warm" case it
will not even be possible to begin to tackle the problem until the form
of the state is explicitly known, and even then it might reasonably be
feared that the ensuing non-linear form of the functional form of the
expression given by (\ref{eq:3.19}) would make it hopelessly
intractible.  What will be shown in the next section is that in the
``cool" limit as described by the equation of state obtained in the
preceeding work\cite{2}, the particular form of the function given by
side of (\ref{eq:3.19}) will be such that the solution will, rather
surprisingly, be obtainable in an explicit analytic form.

\section{The explicit solution for the cool limit case.}
\label{sec:4}

In the ``cool" limit for which the entropy current is describable
simply as a gas of phonons, the corresponding limit form of the equation
of state that is obtained\cite{2} as the relativistic generalisation
of the corresponding classical formula of Landau\cite{9}
is given by an expression of the form
\begin{equation}
{\cal L}= P-3\psi\ ,\label{eq:4.1}
\end{equation}
in which $P$ is the zero temperature pressure function depending only
on the single variable $\mu$ and determining the corresponding zero
temperature sound speed $c_{_{\rm S}}$ by the formula
\begin{equation}
{c^2\over c_{_{\rm S}}^{\,2}}=\mu{d\mu\over dP}{d^2 P\over d\mu^2}\ ,
\label{eq:4.2}
\end{equation}
and $\psi$ is the generalised pressure function of the phonon gas,
which is expressible as a funcion of all three of the independent scalar
variables $s$, $y$, and $\mu$ by the formula
\begin{equation}
\psi = {\tilde\hbar\over 3} c_{_{\rm S}}^{-1/3}
\Big( c^2 s^2+(c_{_{\rm S}}^{\,2}-c^2){y^4\over \mu^2}\Big)^{2/3}\
,\label{eq:4.3}
\end{equation}
where $\tilde\hbar$ is identifiable with sufficient accuracy for most
purposes with the usual Dirac Plank constant $\hbar$, its exact value
being given by
\begin{equation}
\tilde\hbar={9\over 4\pi}\Big({5\pi\over 6}\Big)^{1/3}\hbar\simeq
0.99\hbar\ .\label{eq:4.4}
\end{equation}

It can be seen to follow immediately, just from the fact that this function
$\psi$ is homogeneous, of order $4/3$ in the variables $s$ and $y^2$, that
the function given by (\ref{eq:3.18}) will work out in this ``cool"
limit case simply as
\begin{equation}
\Psi-{\cal L}=4\psi \ .\label{eq:4.5}
\end{equation}
Using (\ref{eq:3.16}), it can be seen that the equation obtained by
substituting (\ref{eq:4.5}) in (\ref{eq:3.18}) can be solved explicitly
to give the formula
\begin{equation}
y={\bar E^{1/2}\over c}\Big({3\bar\Theta\over 4\tilde\hbar}\Big)^{3/2}
c_{_{\rm S}}^{1/2}\Big( {c^2\over {\Gamma}^2}+
{(c_{_{\rm S}}^{\,2}-c^2)\bar E^2\over c^4\mu^2}\Big)^{-1}\ ,\label{eq:4.6}
\end{equation}
whose right hand side is interpretable as a function just of the pair
of variables $\Gamma$ and $\mu$, since the latter determines
$c_{_{\rm S}}^{\,2}$ by the relation (\ref{eq:4.2}) that is obtained
from the zero temperature equation of state.  The formula (\ref{eq:4.6})
thereby gives $y$ directly as an explicit function of the radius $r$
since both $\Gamma$ and $\mu$ are already known by (\ref{eq:3.15}) and
(\ref{eq:3.17}) as functions of $r$.  Having thus evaluated $y$ one can
then immediately obtain the remaining state variable $s$ as a function
of $r$ using (\ref{eq:3.15}) and (\ref{eq:3.16}), thereby completing the
solution of the vortex problem.

The simplest kind of zero temperature equation of state that can be envisaged
for illustrating the application of this formula is the polytropic kind which,
for a given ``rest" mass $m$ per particle, takes the form $P\propto
(\mu-m)^{1+\alpha}$ where the index $\alpha$ is a positive constant (taking
the value $\alpha=3$ in the standard case of a relativistic gas with
kinetic energy large compared with the rest mass energy), so that, by
(\ref{eq:4.2}), one obtains $c_{_{\rm S}}^{\,2}= (c^2/\alpha)(1-m/\mu)$,
which conveniently reduces to a constant in the high density limit for
which $\mu$ becomes large compared with $m$.

\section{The Natural Cut off Radius for a Vortex Cell.}
\label{sec:5}

The validity of the solutions obtained in the previous sections is of course
limited to a finite range of the radial coordinate $r$ which cannot exceed
the critical null corotation radius at which the rigid rotation velocity
reaches the speed of light so that the Lorentz factor $\Gamma$ defined by
(\ref{eq:3.15}) becomes infinite.  There is no danger of approaching
such a singularity in the usual laboratory experiments on exactly
cylindrical superfluid vortices for which wall of the container limits
the radius to a value that is very small compared with the value.  The
occurrence of such a singularity is also avoided in more natural contexts
(such as that of the neutron star material whose analysis is the ultimate
purpose of the present work) under conditions such that the local
superfluid flow can be represented by a honeycomb lattice of hexagonal
vortex cells.  In such an application a cylindrically symmetric solution
of the kind obtained in the previous sections can be used as approximate
description of the motion within an individual hexagonal vortex cell.
In a stationary state the honeycomb lattice will be in a state of rigid
rotation with an angular velocity $\Omega$ that must be the same as that
of the normal constituent (if any) in order to avoid the presence of
dissipation (which would be incompatible with strict stationarity).  In
such a configuration, each approximately cylindrical vortex cell will
have a natural outer cut off radius,
\begin{equation}
r=\bar r\ \label{eq:5.1}
\end{equation}
say, where $\bar r$ is just {\it half} the distance between neighbouring
vortex cores.  (The axial symmetry approximation should be extremely
accurate in the inner region $r<<\bar r$ where most of the vortex energy
is concentrated, so the errors due to neglect of the breakdown from
axial to hexagonal symmetry in the outer regions should be relatively
small.) The symmetry between neigbouring vortices of the lattice implies
that at the midpoint between nearest neighbour vortex cores the angular
velocity, not just of the normal flux $s^\mu$, but also of the particle
flux $n^\mu$ should should agree with that angular velocity of the
lattice, with respect to the frame in which the longitudinal velocity
$V$ of the rigid normal flow vanishes.  This means that provided we fix
the choice of the longitudinal Killing vectors $k^\mu$ and $\ell^\mu$ by
imposing the requirement
\begin{equation}
V=0\label{eq:5.2}
\end{equation}
then the appropriate cut off radius (\ref{eq:5.1}) can be characterised
simply by the condition
\begin{equation}
\omega=\Omega\ ,\label{eq:5.3}
\end{equation}
where $\omega$ is the radially variable angular velocity of the particle
current as introduced in (\ref{eq:3.1}). Since it can be seen from
(\ref{eq:2.11}) and (\ref{eq:3.12})
that this angular velocity variable will be given by
\begin{equation}
{\omega r^2\over c^2}={M-\Omega r^2{\cal A}\sigma\over
E-c^2{\cal A}\sigma}\ ,\label{eq:5.4}
\end{equation}
the equation (\ref{eq:5.3}) is easily soluble.  The natural cut off
radius value is thereby found to be given by the simple formula
\begin{equation}
\bar r^2={c^2 M\over E\Omega}\ .\label{eq:5.5}
\end{equation}
This formula can be used to relate the angular velocity $\omega$ of the
rigidly rotating frame to the corresponding circumferential value $\bar
w$ of of the circulation $w$ per unit area.  Since the momentum
circulation $\kappa$ say will be given in terms of the angular momentum
constant $M$ by $\kappa=2\pi M$, the corresponding mean vorticity, $w$
for a circuit of radius $r$ will be given by
\begin{equation}
w ={2M\over r^2}\ ,\label{eq:5.6}
\end{equation}
so that for the corresponding effective macroscopic vorticity, as given
by the momentum circulation per unit area for each entire vortex cell, one
obtains the formula
\begin{equation}
\bar w = 2E\Omega/c^2 \ .\label{eq:5.7}
\end{equation}

For the purpose of applications to the analysis of macroscopic properties of
a vortex lattice, it is of particular interest to evaluate total integrated
quantities, such as the total effective longitudinal energy momentum tensor
of a vortex as functions of the corresponding total longitudinal currents,
within the relevant cut off radius (\ref{eq:5.1}).

It is shown in the appendix how the asymptotic deviation of the sectionally
averaged value $\overline Q$ of any suitable physical quantity $Q$ from
its value $Q_{_{\ominus}}$ in an appropriately chosen homogeneous
reference state goes like $r^{-2}\ln{r}$ for large cut-off distances
$r$, the multiplicative factor depending on a corresponding ``net
asymptotic deviation coefficient" $\widehat Q$.  In order for the
averaging process to be meaningful, it is not necessary that the physical
quantity $Q$ under consideration should be a scalar in the 4-dimensional
sense, but is sufficient that it should be a scalar with respect to the
coordinates $\phi$ and $r$ of the section at constant $t$ and $z$ over
which the integrals are taken, as is the case for tensor components as
longitudinally projected onto the symmetry axis whose coordinates are
the subset $\{x^{_{i}}\}\leftrightarrow\{t,z\}$, for $\{i\}=\{0,1\}$
within the full set $\{x^\mu\}$ introduced in (\ref{eq:3.10}). This applies
in particular to the corresponding longitudinally projected components
of the stress tensor, whose sectionally integrated total will be given
in terms of the corresponding averaged values $\overline{T^j_{\ k}}$ by
\begin{equation}
\langle T^j_{\ k}\rangle=\pi r^2\overline{T^j_{\ k}}\ .\label{eq:5.8}
\end{equation}
The foregoing formulae are also valid for the trace of the orthogonally
projected part of the stress tensor, which is twice what is known as the the
{\it lateral} pressure ${\mit\Pi}$ say, whose definition is expressible
equivalently by
\begin{equation}
2{\mit\Pi}=T^\rho_{\ \rho}-T^i_{\ i}\ ,\label{eq:5.9}
\end{equation}
so that its average will be expressible in terms of the scalar functions
introduced in Section 1 by
\begin{equation}
2\overline{\mit\Pi}=3\overline\Psi-\overline \Lambda-\overline{T^i_{\ i}} .
\label{eq:5.10}
\end{equation}

According to the general formula obtained in the appendix, the averages in
(\ref{eq:5.8}) will be given asymptotically by an expression of the form
\begin{equation}
\overline{T^{j}_{\ k}}- T^{\,j}_{_{\ominus}k}\sim\widehat{T^{j}_{\ k}}
{Mw\over 4c^2\mu^2}\,
{\rm ln}\Big\{ {w_{\!_\odot}\over w}
\Big\} , \label{eq:5.11}
\end{equation}
as $w \rightarrow 0$ (using the standard notation convention according to
which the symbol $\sim$ relates quantities whose ratio tends to unity in
the limit under consideration) where $T^{\,j}_{_{\ominus}k}$ are the
corresponding reference state values, and $w_{\!_\odot}$ is a fixed
vorticity value determined according to (\ref{eq:5.6}) by a suitably
chosen fixed ``sheath" radius $r_{\!_\odot}$ whose precise specification
is unimportant when $w$ is sufficiently small.

\section{Stress - Energy coefficients for a Cool Vortex Cell.}
\label{sec:6}

The purpose of this  section is to use the procedure described in the
appendix to evaluate the coefficient $\widehat{T^{j}_{\ k}}$, taking the
variables $q$ that fix the reference state to be simply the longitudinal
components $\mu^i$ of the particle momentum (the other components being
irrelevant since the axial symmetry ensures that they average to zero ) and
the longitudinal components $s^i$ of the entropy current. This choice for
the reference state is not at all compelling, and we shall consider for
instance in the next section, in the  cold limit, a reference state defined
by the longitudinal components of the particle current. Our choice here is
simply motivated by the computational simplicity when one will wish to apply
the sectional averaging to a vortex solution, for which the most convenient
variables are the particle momentum and the entropy current as illustrated
in Section~\ref{sec:3}.

The reference state is thus chosen to satisfy the condition that the
corresponding reference momentum components $\mu_{i{_\ominus}}$ and reference
current components $s^i_{_\ominus}$ agree respectively with the average
momentum components $\overline{\mu_i}$ and the current components
$\overline{s^i}$ of
the vortex, so that in the notation of (\ref{eq:A9}) one has
\begin{equation}
\delta_{_\ominus}\overline{\mu_i} =0,\qquad
  \delta_{_\ominus}\overline{s^i}=0\ . \label{eq:6.1}
\end{equation}
According to (\ref{eq:A23}), this means that the required ``net" asymptotic
deviation coefficients $\widehat{T^{j}_{\ k}}$ will be given in terms of
the corresponding ``gross'' asymptotic deviation coefficients
$\widetilde{T^{j}_{\ k}}$ by
\begin{equation}
  \widehat{T^{j}_{\ k}}=\widetilde {T^{j}_{\ k}}
-\widetilde{s^i}\,{\partial\over\partial s^i}{T^{j}_{\ k}} \ ,\label{eq:6.2}
\end{equation}
using the fact that
\begin{equation}
\widetilde{\mu_i}=0,\label{eq:6.3}
\end{equation}
which follows obviously from (\ref{eq:3.12}).

To apply the formula (\ref{eq:6.2}) we first need to use (\ref{eq:A19})
to obtain the required ``gross" asymptotic deviation coefficients
$\widetilde{s^i}$ for the entropy current itself, whose ``net" asymptotic
deviation coefficients will automatically vanish $\widehat{s^i}= 0$ as
an automatic consequence of the choice (\ref{eq:6.1}). In the case of the
vortex solution, the dependence of $s^i$ on $\mu$ is completely confined
to the coefficient $\sigma$ as can be seen from (\ref{eq:3.12}), which
is itself related to the ``cross" scalar $y^2$ according to
(\ref{eq:3.13}), so the deviation formula (\ref{eq:A19}) gives
\begin{equation}
\widetilde{s^i}=-{\mu\over y^2}{dy^2\over d\mu}s^i \ ,\label{eq:6.4}
\end{equation}
where the limit $r\rightarrow\infty$ is implicit in
the right hand side, as will be the case  in all the
formulas giving asymptotic coefficients.

We are now ready to consider the application  to the stress tensor which
can be written in the form
\begin{equation}
T^\rho_{\ \sigma}=\Phi^2\left(\mu^\rho\mu_\sigma+{\cal K} s^\rho s_\sigma
\right)
+\left[{\cal L}+c^2\Phi^2 \left(s^2 {\cal K}+y^2{\cal A}\right)\right]
g^\rho_{\ \sigma}\ .\label{eq:6.5}
\end{equation}

It is now straightforward to compute $\widetilde{T^j_{\ k}}$ by using
systematically the definition (\ref{eq:A19}). We skip here the details of
the calculation, inviting the reader to consult the similar but simpler
procedure for the cold limit that we will detail in the next section.
The result involves many terms. It turns out that when one adds to the
expression for $\widetilde{T^j_{\ k}}$ the extra term in order to get
the net asymptotic coefficient $\widehat{T^j_{\ k}}$, a lot of
cancellations occur, some due to the relation
\begin{equation}
 ds={s\over y^2}dy^2+{1\over 2}s\Gamma^2 d\left(\Gamma^{-2}\right)
\ ,\label{eq:6.6}
\end{equation}
so that the final result can be simply expressed as
\begin{equation}
\widehat{T^j_{\ k}}=-\mu{\partial \Phi^2\over \partial\mu}\mu^j\mu_k
+{\mu^2\over s}{\partial\Phi^2\over\partial s}s^js_k+\mu^2 c^2\left(-\Phi^2
+s{\partial \Phi^2\over\partial s} + y^2{\partial \Phi^2
\over\partial y^2}\right) g^j_{\ k}\ .   \label{eq:6.7}
\end{equation}

We now consider the asymptotic behaviour of the lateral pressure contribution
(\ref{eq:5.10}) which can be seen to be given locally by
\begin{equation}
{\mit\Pi}={\cal L}+c^2\Phi^2\left[s^2 {\cal K}+ y^2 {\cal A}
+{1\over 2}\big(\mu^2_{_\infty}-\mu^2\big)
+{1\over 2}{\cal K}\sigma^2\big(\Gamma^{-2}_{_\infty}-\Gamma^{-2}\big)
\right]\ .\label{eq:6.8}
\end{equation}
Although the expression that one obtains ``naively" for $\widetilde{\mit\Pi}$,
i.e. by just applying the definition (\ref{eq:A16}), involves a lot of terms,
it
can be checked, by using (\ref{eq:6.6}) as well as the differential of the
relation (\ref{eq:3.18}) written in the form
\begin{equation}
{s^2\over y^2}\Phi^2{\cal K} +\Phi^2{\cal A}= {\bar\Theta\over\overline E}
 \ ,\label{eq:6.9}
\end{equation}
whose right hand side is constant, that the ``gross"
asymptotic deviation coefficient cancels out altogether, i.e.  one
simply gets
\begin{equation}
\widetilde{\mit\Pi}=0\ .\label{eq:6.10}
\end{equation}
This result  could of course have been predicted in advance from the
requirement of overall stress balance on the outer boundary.

Despite of the stress balance condition (\ref{eq:6.10}), the lateral
pressure will nevertheless have a non vanishing ``net" asymptotic
deviation coefficient that will be given by
\begin{equation}
\widehat{\mit\Pi}=\mu \left(s{\partial \Phi^2\over\partial s}
+y^2{\partial \Phi^2\over \partial y^2}\right).\label{eq:6.11}
\end{equation}

The preceeding results are valid for any equation of state ${\cal L}$. We
now specialize these results to the ``cool'' equation of state (\ref{eq:4.1})
for which we have explicit vortex solutions.  In this case, the dilation
coefficient is given by
\begin{equation}
\Phi^2={1\over\mu c^2}{dP\over d\mu}-{3\over\mu c^2}{\partial\psi\over
\partial \mu},\label{eq:6.12}
\end{equation}
the second term on the right hand side being homogeneous of order $4/3$
in the variables $s$ and $y^2$.  It can thus be checked easily that the
lateral pressure is simply given by
\begin{equation}
\widehat{\mit\Pi}=-{4\over c^2}{\partial\psi\over\partial\mu}\ .\label{eq:6.13}
\end{equation}
whereas the coefficient $\widehat{T^j_{\ k}}$ is given by substituting
(\ref{eq:6.12}) in (\ref{eq:6.7}).

\section{Stress - Energy coefficients for a Cold Vortex Cell.}
\label{sec:7}

In this final section, we restrict our attention to the case of the cold
limit in which there is no entropy vector.  The superfluid in this zero
temperature limit reduces to a particular case of perfect fluid and the
Lagrangian ${\cal L}$ reduces simply to the pressure function $P(\mu)$.
In this simple case the current will by given, in terms of the ``dilatonic"
amplitude $\Phi$ that plays a key role in the vorticity variational
formulation\cite{10} of the zero temperature fluid model, just by
\begin{equation}
n^\rho=\Phi^2\mu^\rho\ , \label{eq:7.1}
\end{equation}
where $\Phi^2$ and its derivative are given by
\begin{equation}
\Phi^2={n\over\mu}\ , \hskip 1 cm \mu{d\Phi^2\over d\mu}
=\Phi^2 \Big({c^2\over c_{_{\rm S}}^2}-1\Big) \ .\label{eq:7.2}
\end{equation}
and
where the particle number density $n$ is defined by
\begin{equation}
n^\rho n_\rho=-c^2n^2\label{eq:7.3}
\end{equation}
and can also be derived from the pressure function $P(\mu)$ by
\begin{equation}
n=c^{-2}{dP\over d\mu}\ .\label{eq:7.4}
\end{equation}
One can also introduce the energy density of the fluid, $\rho(n)$,
as the Legendre transform of the pressure function $P(\mu)$,
\begin{equation}
\rho(n)=\mu n -c^{-2}P(\mu)\ . \label{eq:7.5}
\end{equation}
It follows from (\ref{eq:3.12}) that the longitudinal and angular velocities
$v$ and $\omega$ introduced in (\ref{eq:3.1}) and the corresponding
amplitude factor $\nu$ will be given by
\begin{equation}
v={c^2L\over E}\ , \qquad \omega={c^2 M\over E r^2}\ ,\hskip 1 cm
\nu= {E\over c^2}\Phi^2 .\label{eq:7.6}
\end{equation}
Finally, in
this zero temperature limit, the stress tensor  has the simple
perfect fluid form
\begin{equation}
  T^\rho_{\ \sigma}=n^\rho\mu_\sigma+Pg^\rho_{\ \sigma}\ .\label{eq:7.7}
\end{equation}

It is evident from (\ref{eq:6.2}) that in the cold limit for which the
entropy current vanishes the ``gross" deviations (with respect to the
large distance limit) are the same as the ``net" deviations with respect
to a reference state of the kind postulated in the preceeding section.
This is because the cold limit of such a reference state limit is
characterised just by the longitudinal momentum components, which are
radially uniform, so that deviations from their large distance limit
values simply vanish:
\begin{equation}
\delta_{_\infty}\overline{\mu_i} =0\ . \label{eq:7.8}
\end{equation}
Using (\ref{eq:7.2}) the ``gross" variation coefficients for the longitudinal
stress energy tensor in the cold limit can therefore be obtained simply
by setting the entropy current $s^i$ to zero in (\ref{eq:6.7}) which gives
\begin{equation}
\widetilde{T^j_{\ k}}=\left(1-{c^2\over c_{_{\rm S}}^2}\right) n^j\mu_k-(\rho
c^2+P) g^j_{k},
\label{eq:7.9}
\end{equation}
while by (\ref{eq:6.9}) the corresponding ``gross" lateral pressure simply
vanishes.

Introducing the longitudinal unit flow vector, $u^\rho$ say, and the
corresponding spacial projection tensor
\begin{equation}
  \gamma^\rho_{\sigma}=g^\rho_{\ \sigma}+{1\over c^2}u^\rho u_\sigma
\ ,\qquad\quad u^\rho u_\rho=-c^2,\label{eq:7.10}
\end{equation}
according to the component prescription
$\mu_{_\ominus}\{u^{_0},u^{_1},u^{_2},u^{_3}\}=$
$\{\overline{\mu^{_0}},\overline{\mu^{_1}},0,0\} $ (in the coordinate
system introduced in (\ref{eq:3.10}) that we have been using) so that
the uniform asymptotic limit state is characterised by the condition
that its momentum should simply have the form
\begin{equation}
\mu_{_{\ominus}\rho}=\mu_{_\ominus} u_\rho\ ,\label{eq:7.11}
\end{equation}
the meaning of the formula (\ref{eq:7.9}) can be interpreted as follows.
In terms of the corresponding pressure $P_{_\ominus}$ and energy density
$\rho_{_\ominus}$ given, via the equation of state (\ref{eq:2.1}) by
\begin{equation}
P_{_\ominus}=P\{\mu_{_\ominus}\} \ ,\qquad\quad \rho_{_\ominus}=
\mu_{_\ominus}{dP\over d\mu}_{_\ominus}-P_{_\ominus} \ ,\label{eq:7.12}
\end{equation}
the stress energy tensor of the uniform asymptotic limit state will
be given by
\begin{equation}
  T^{\,\rho}_{_{\ominus}\sigma}=\rho_{_\ominus} u^\rho
    u_\sigma+P_{_\ominus}\gamma^\rho_{\ \sigma} \ .\label{eq:7.13}
\end{equation}
The significance of (\ref{eq:7.9}) is that that it specifies the coefficient
in the asymptotic formula
\begin{equation}
  \langle{T^{j}_{\ k}}\rangle-\langle T^{\,j}_{_{\ominus}k}\rangle \sim
{\pi M^2\over 2 c^2 \mu^2} \widetilde{T^{j}_{\ k}}\,\ln
\Big\{ {w_{\!_\odot}\over w} \Big\}  \label{eq:7.14}
\end{equation}
for the difference of the total sectionally integrated stress momentum
energy density components from the values they would have for the uniform
asymptotic limit state (if the vortex were absent) as a function of the cut
off radius $r$ as given in terms of the corresponding circulation per unit
area $w$ by (\ref{eq:5.6}). The components $T^{\,j}_{_{\infty}k}$ in this
expression are just the longitudinal subset of the 4-dimensional set
given by (\ref{eq:7.13}) while the formula (\ref{eq:7.9}) for the
coefficients $\widetilde{T^{j}_{\ k}}$ can be rewritten equivalently, in
a form more directly comparable with (\ref{eq:6.13}), as
\begin{equation}
\widetilde{T^{j}_{\ k}}=(\rho c^2+P) \Big( (2c^{-2}-c_{_{\rm S}}^{-2})
u^j u_k -\gamma^j_{_k}\Big) .\label{eq:7.15}
\end{equation}
Substituting this in (\ref{eq:7.14}) gives the simple explicit formula
\begin{equation}
\langle{T^{j}_{\ k}}\rangle-\langle T^{\,j}_{_{\ominus}k}\rangle
\sim {\pi M^2\over 2}\Phi^2
\Big( (2c^{-2}-c_{_{\rm S}}^{-2}) u^j u_k -\gamma^j_{_k}\Big)\,
\ln \Big\{ {w_{\!_\odot}\over w} \Big\}\ . \label{eq:7.16}
\end{equation}

The final formula (\ref{eq:7.16}) is interpretable as meaning that relative
to the
reference state labelled by $_{\ominus}$,  with the same longitudinal
momentum component values $\overline\mu_i$, the vortex cell has an effective
energy per unit length, $\big[ U\big]_{_\ominus}$ say, that is given by
\begin{equation}
 \big[ U\big]_{_\ominus}\sim \left(2-{c^2\over c_{_{\rm S}}^2}\right)
\big[ T\big]_{_\ominus}\label{eq:7.17}
\end{equation}
where, $\big[ T\big]_{_\ominus}$ is the corresponding effective tension, which,
for the usual case with the lowest quantum value, $M=\hbar$, will be given
as a function of the  mean vorticity $w$ defined by (\ref{eq:5.6}) and the
amplitude $\Phi$ defined by (\ref{eq:6.2}) as
\begin{equation}
\big[ T\big]_{_\ominus}\sim {\pi \hbar^2\over 2}\Phi^2 \, \ln
\Big\{ {w_{\!_\odot}\over w}\Big\} \ ,\label{eq:7.18}
\end{equation}
wherein --~in view of the fact that higher order corrections have been
neglected~-- it is sufficiently accurate to take the quantities involved
to be those characterising the simple vortex free reference state.

In the non-relativistic limit, in which one can make the substitution
$\Phi^2\sim\rho/m^2$ where $m$ is the relevant fixed Newtonian mass per
particle, the expression (\ref{eq:7.18}) can be seen to agree with the
well known formula given by Hall\cite{11} for the effective vortex
tension defined as the integrated longitudinal pressure deficit relative
to the asymptotic value.  Hall pointed out that this value is the same
as that given by the Feynman formula for the energy per unit length, as
evaluated in the incompressible case.  It is to be remarked that the
incompressible Newtonian case corresponds to the non-relativistic limit
of the ``stiff" case characterised by $c_{_{\rm S}}^2=c^2$, and that for
a``softer" material the energy per unit length given by (\ref{eq:7.16})
will be relatively diminished, so much so that it will actually be
negative whenever $c_{_{\rm S}}^2<c^2/2$, as will typically be the case
in realistic applications, including all except perhaps the most central
regions of neutron stars.  This energy deficit is simply due to the
central matter deficit resulting, unless the material is sufficiently
stiff, from the centrifugal effect. It will be seen in the next section
that, when normalised with respect to a fixed amount of conserved matter,
the energy per unit length will always be positive, and furthermore that
its value will remain equal to that given by (\ref{eq:7.18}) even for
material that is quite ``soft".

\section{Stress - Energy with respect to the Current Flux Reference State.}
\label{sec:8}

Rather than the prescription of Section~\ref{sec:6}, which as we have
seen would give no difference between ``gross" and ``net" deviations in
the cold limit, let proceed with the analysis of the cold limit on the
base of a prescription that is less mathematically trivial.  Instead of
merely choosing the reference state to be specified by the longitudinal
components of the momentum covector, it is more useful for many physical
purposes to choose a reference state to be fixed by the longitudinal
components of the particle current flux (the other components being
irrelevant since the axial symmetry ensures that they average to zero).
To avoid confusion with the previous choice of the reference state that
was labelled by the symbol $_{\ominus}$, we shall use a slightly
modified symbol, $_\oslash$, to label this new reference state which is
chosen so as to satisfy the condition that the corresponding reference
current components $n^i_{_{\oslash}}$ agree with average current
components $\overline{n^i}$ of the vortex, so that in the notation of
(\ref{eq:A8}) one has
\begin{equation}
\delta_{_{\oslash}}\overline{n^i} =0\ . \label{eq:8.1}
\end{equation}

According to the appendix, in the same way as in the expression
(\ref{eq:6.2}), the relation between the ``net" and ``gross'' asymptotic
deviation coefficients will be given in this case by
\begin{equation}
\widehat{T^{j}_{\ k}}=\widetilde {T^{j}_{\ k}}
-\widetilde{n^i}\,{\partial\over\partial n^i}{T^{j}_{\ k}} \ .\label{eq:8.2}
\end{equation}

To use this formula we  need to  the  ``gross" asymptotic deviation
coefficients $\widetilde{n^i}$ for the current itself. Using the last of the
expressions (\ref{eq:7.6}), it can be seen that the asymptotic deviation
formula (\ref{eq:A16}) will give
\begin{equation}
 \widetilde{n^i}=\Big(1-{c^2\over c_{_{\rm S}}^2}\Big)n^i \ ,\label{eq:8.3}
\end{equation}
in which the ratio of the light and sound speeds is given by (\ref{eq:4.2}).

In view of the fact that, by (\ref{eq:3.12}), the relevant components of
the momentum covector $\mu_\rho$ are constant, as also are those of the
metric, they can be taken outside the averaging process so as to allow
us to write
\begin{equation}
\widetilde{T^j_{\ k}}=\widetilde{n^j}\mu_k +\widetilde P g^j_{_k}\ .
\label{eq:8.4}
\end{equation}
Therefore to complete the evaluation of $\widetilde{T^j_{\ k}}$ all that
remains
to be done is to obtain the asymptotic deviation coefficient of the
pressure scalar $P$. Applying (\ref{eq:A16}) and using (\ref{eq:7.4}) and
(\ref{eq:7.5}) it can be seen that the required result is given simply by
\begin{equation}
  \widetilde P=-(\rho c^2+P) \ .\label{eq:8.5}
\end{equation}
Putting this together with (\ref{eq:8.3}) in (\ref{eq:8.4}) we obtain a
direct derivation of the complete expression (\ref{eq:7.9}) for the ``gross"
asymptotic deviation coefficient of the stress momentum energy density tensor.

Proceeding towards the evaluation of the ``net" deviation coefficients in
which we are ultimately interested, the next step is to use (\ref{eq:8.3})
again for working out the second term required for the application of
(\ref{eq:8.2}).  The contribution from the pressure scalar $P$, which
(in view of the asymptotic agreement ${\mit\Pi}\sim P$ as
$\mu\rightarrow\mu_{_\infty}$) is the same as that for ${\mit\Pi}$, is
given simply by
\begin{equation}
 {\partial P\over\partial n^i}=-{c_{_{\rm S}}^2\over c^2}\mu_i \ ,
  \Rightarrow \widetilde{n^i}\,{\partial P\over\partial n^i} =
  \widetilde{n^i}\,{\partial{\mit\Pi}\over\partial n^i}
  =\Big({c_{_{\rm S}}^2\over c^2}-1\Big) (\rho c^2+P)\ .\label{eq:8.6}
\end{equation}
As an immediate corollary we see from (\ref{eq:8.2}) that in to contrast
to the case where the reference state is specified by the momentum, the
lateral pressure will have a non vanishing ``net" asymptotic deviation
coefficient that will be given by
\begin{equation}
 \widehat{\mit\Pi}=\Big(1-{c_{_{\rm S}}^2\over c^2}\Big) (\rho c^2+P)\
     .\label{eq:8.7}
\end{equation}

After analogously evaluating the other relevant contributions their
combination in the complete longitudinally projected stress tensor is
obtained as
\begin{equation}
 \widetilde{n^i}\,{\partial\over\partial n^i} T^j_{\ k}
=\Big({c_{_{\rm S}}^2\over c^2} -{c^2\over c_{_{\rm S}}^2}\Big)n^j\mu_k
 +\Big({c_{_{\rm S}}^2\over c^2}-1\Big)(\rho c^2 +P)g^j_{\ k}\ .\label{eq:8.8}
\end{equation}
When this last, rather unweildy, expression is subtracted off from
(\ref{eq:7.9}) in accordance with the prescription (\ref{eq:8.2}), we
end up with a comparitively simple expression for the required ``net"
deviation coefficient of the stress tensor, which is expressible just by
\begin{equation}
  \widehat{T^j_{\ k}}=\Big(1-{c_{_{\rm S}}^2\over c^2}\Big)n^j\mu_j
-{c_{_{\rm S}}^2\over c^2}(\rho c^2 +P)g^j_{\ k}\ ,\label{eq:8.9}
\end{equation}
in which, as in preceeding formulae, the relevant values of the quantities
involved are those of the large distance limit.

This result is to be interpreted with respect to the reference state as
characterised by a density  $n_{_{\oslash}}$ given in terms of the average
flow by the component prescription
$n_{_{\oslash}}\{u^{_0},u^{_1},u^{_2},u^{_3}\}=$
$\{\overline{n^{_0}},\overline{n^{_1}},0,0\}\ $ where $u^\rho$ is the
same unit vector as was introduced in (7.10) so that the corresponding
current has the form
\begin{equation}
  n_{_{\oslash}}^\rho=n_{_{\oslash}} u^\rho\ . \label{eq:8.9*}
\end{equation}
In terms of the corresponding density $\rho_{_{\oslash}}$ and pressure
$P_{_{\oslash}}$  given, via the equation of state (\ref{eq:7.5}), by
\begin{equation}
\rho_{_{\oslash}}=\rho\{n_{_{\oslash} }\}\ ,\qquad\quad
P_{_{\oslash}}=n_{_{\oslash}}{d\rho\over d n}_{_{\oslash}}
-\rho_{_{\oslash}} \ ,\label{eq:8.10}
\end{equation}
The stress energy tensor of the uniform asymptotic limit state will
be given in terms of these quantities an expression of the perfect fluid
form that is the formal analogue of (\ref{eq:7.13}) namely
\begin{equation}
  T^{\,\rho}_{_{\oslash}\sigma}=\rho_{_{\oslash}} u^\rho
u_\sigma+P_{_{\oslash}}\gamma^\rho_{\ \sigma} \ .\label{eq:8.11}
\end{equation}
With respect to this reference state, the  expression for the total
sectionally integrated stress tensor will be given by
\begin{equation}
  \langle{T^{j}_{\ k}}\rangle-\langle T^{\,j}_{_{\oslash}k}\rangle \sim
{\pi M^2\over 2}\Phi^2
\Big({1\over c^2} u^j u_k -{c_{_{\rm S}}^2\over c^2}\gamma^j_{_k}\Big)
\,\ln \Big\{ {w_{\!_\odot}\over w} \Big\} . \label{eq:8.12}
\end{equation}

This last formula  is  interpretable as meaning that relative to the
reference state labelled $_{\oslash}$, with the same average longitudinal
current vector $\overline n^i$, the  vortex cell has an effective
tension, $\big[ T\big]_{_{\oslash}}$ say, and an effective energy density per
unit length, $\big[ U\big]_{_{\oslash}}$ say, that are related by
\begin{equation}
  \big[ T\big]_{_{\oslash}} \sim {c_{_{\rm S}}^2\over c^2} \big[ U
    \big]_{_{\oslash}} \ ,\label{eq:8.13}
\end{equation}
where $\big[ U\big]_{_{\oslash}}$  is given by
\begin{equation}
  \big[ U\big]_{_{\oslash}} \sim {\pi \hbar^2\over 2}\Phi^2 \,{\rm ln}
\Big\{ {w_{\!_{ {\!_\odot} }}\over w}\Big\} \ ,\label{eq:8.14}
\end{equation}
for the lowest quantum value $M=\hbar$.

It is to be remarked that relative to the reference state used in the
present section, as defined in terms of a fixed quantity of conserved matter,
the effective tension can be seen from (\ref{eq:8.13}) to tend to zero
in the limit $c_{_{\rm S}}^2<< c^2$.  It is also to be noted that --~as
a generalisation of the non-relativistic Hall\cite{11} equality~-- the last
formula (\ref{eq:8.14}) for the energy per unit length is the same as the
formula (\ref{eq:7.18}) that was obtained for the effective tension,
albeit relative to a different reference state.  However it is only in
the ``stiff" case characterised by $c_{_{\rm S}}^2=c^2$ that, as in the
incompressible non-relativistic case, the effective tension will be
equal to the energy per unit length with respect to the {\it same}
reference state.

\appendix
\section*{Asymptotic forms of Sectional integrals and averages.}

The purpose of this appendix is to consider sectional integrals and averages
of a generic local radially dependent physical quantity $Q$ in the
asymptotic limit as the outer cut off radius ${\bar r}$ of the section
tends to infinity, i.e.  in the limit as the mean vorticity,
\begin{equation}
\bar w ={2M\over {\bar r}^2},\label{eq:A1}
\end{equation}
tends to zero. We use angle brackets to denote the sectional integral
of $Q$ over the entire two dimensional section $\Sigma$ say out to the
exterior cut off radius ${\bar r}$ of the cylindrically symmetric vortex
region under consideration, i.e.  we write
\begin{equation}
\langle Q\rangle =\int_\Sigma Q\, d\Sigma=2\pi\int^{\bar r} Q\, r dr\ .
\label{eq:A2}
\end{equation}
Such an integral will typically diverge in the asymptotic limit
$\bar r\rightarrow\infty$, unlike the corresponding average,
\begin{equation}
  \overline Q={\langle Q\rangle\over \pi \bar r^2}={2\over\bar r^2}
  \int^{\bar r} Q\, r dr . \label{eq:A3}
\end{equation}
which will typically tend to a well defined finite limit.
Provided that $Q$ itself  is asymptotically well behaved in the sense
of tending smoothly to a well defined asymptotic limit
\begin{equation}
   Q\rightarrow Q_{_\infty} \label{eq:A4}
\end{equation}
as $r\rightarrow\infty$, its average will obviously converge to the same
asymptotic limit i.e. we shall have
\begin{equation}
\overline Q\rightarrow Q_{_\infty} \ .\label{eq:A5}
\end{equation}
Our interest will therefore focus for the time being on the non trivial
difference between the average and its limit, as defined by
\begin{equation}
  \delta_{_\infty}\overline Q=\overline Q- Q_{_\infty} \, \label{eq:A6}
\end{equation}
and on the corresponding difference as defined for the total integral by
\begin{equation}
 \delta_{_\infty}\langle Q\rangle=\langle Q\rangle-\pi \bar r^2 Q_{_\infty}=
 \pi \bar r^2\delta_{_\infty}\overline Q \ .\label{eq:A7}
\end{equation}
Whereas it is evident from (\ref{eq:A5}) that $\delta_{_\infty}\overline Q$
will tend to zero, on the other hand $\delta_{_\infty}\langle Q\rangle$
will still diverge, albeit only logarithmicly, i.e.  much less strongly
than $\langle Q\rangle$, as $\bar r\rightarrow\infty$.

In order to analyse this limit it is useful to decompose the range of
integration into an ``asymptotic'' zone, $\bar r>r>r_{\!_\odot}\ $,
$\bar w<w<w_{\!_\odot}$ and an inner ``sheath" zone, $r<r_{\!_\odot}\ $,
$w>w_{\!_\odot}$, taking the sheath radius $r_{\!_\odot}$ to be large
compared with the central core radius within which the superfluidity
breaks down but small compared with the outer cut off radius $\bar r$.
It can be seen that the average $\overline Q$ as taken over the entire
section out to $\bar r$ will be given in terms of the sheath average
$\overline Q_{\!_\odot}$ as taken over just the inner zone extending
only to $r_{\!_\odot}$ by an expression of the form
\begin{equation}
  \overline Q={\bar w\over w_{\!_\odot}}\overline Q_{\!_\odot}
  +\bar w\int_{\bar w}^{w_{\!_\odot}}Q {dw\over w^2}\ .\label{eq:A8}
\end{equation}

The subject of our ultimate concern  is not so much the total integral or
average as given by the foregoing expressions but rather just the part
thereof that is attributable to the vortex {\it per se}, after subtracting
off a contribution that would have been there in its absence. What we are
specially interested in are not the ``gross"
differences defined by (\ref{eq:A4}) but the corresponding ``net" differences
\begin{equation}
  \delta_{_\ominus}\overline Q=\overline Q-\overline Q_{_{{\ominus}}}\ ,
\label{eq:A9}
\end{equation}
where $Q_{_{\ominus}}$ is the value that the quantity under consideration
would have in some appropriately chosen reference state. We take this
reference state not just vortex free but more specificly {\it uniform},
which of course entails the exact equality $Q_{_{\ominus}}=\overline
Q_{_{\ominus}}$.  In practice we shall compute this ``net'' difference
by starting from the ``gross'' difference and by using the relation
\begin{equation}
  \delta_{_\ominus}\overline Q =\delta_{_\infty}\overline Q
-\delta_{_\infty} Q_{_{\ominus}} \label{eq:A10}
\end{equation}
where $\delta_{_\infty} Q_{_{\ominus}}$ is the difference between
$Q_{_{\ominus}}$ and the aymptotic value $Q_{_\infty}$ in the vortex
state, i.e.
\begin{equation}
  \delta_{_\infty} Q_{_{\ominus}}=\overline Q_{_{\ominus}}
    -Q_{_\infty}\ .\label{eq:A11}
\end{equation}
It is to be noted that the possibility of using a reference state that is
uniform is made possible by our neglect of gravity in the work from the end
of Section~\ref{sec:3} onwards.

The complete specification of the reference state and the corresponding
difference $\delta_{_\infty} Q_{_{\ominus}}$ is not fixed just by the
uniformity requirement but is is a matter of discretion, depending on
the physical context under consideration.  The most natural choice for
many purposes is to fix the reference state by the requirement that that
it give unchanged values for the total integrals and averages of the
independent current variables, but for some purposes other choices might
be more convenient.  Before adopting any particular convention as to the
choice of reference state let us first consider the ``gross" difference
denoted by $\delta_{_\infty} Q$ which is already well defined {\it a
priori}.

In order to see the nature of the dominant contribution to the ``gross"
differences in the large $r$ limit, we exploit the fact that the equation
(\ref{eq:3.17}) for the radial dependence of the chemical potential will be
expressible in the form
\begin{equation}
  \delta_{_\infty}\mu^2=-{M\over 2c^2} w \ .\label{eq:A12}
\end{equation}
Assuming that the quantity $Q(r)$ under consideration has a
dependence on the radial coordinate $r$ only through the intermediary of
either $\mu(r)$ or $\Gamma (r)$, one can use a Taylor expansion for
Q considered as function of the two variables $\mu^2$ and $\Gamma^{-2}$,
\begin{equation}
Q(r)=Q(\bar r)+\left[{dQ\over d\mu^2}\right]_{\bar r}\left(\mu^2(r)
-\mu^2(\bar r)\right)
+\left[{dQ\over d(\Gamma^{-2})}\right]_{\bar r}
\left(\Gamma^{-2}(r)-\Gamma^{-2}(\bar r)\right)+...  \label{eq:A13}
\end{equation}
where we have written explicitly only the first terms of the Taylor
expansion.
Using (\ref{eq:3.12}) and (\ref{eq:3.15}), one gets
\begin{equation}
\mu^2(r)-\mu^2(\bar r)=-{M\over 2c^2}\left(w-\bar w\right),\label{eq:A14}
\end{equation}
and
\begin{equation}
\Gamma^{-2}(r)-\Gamma^{-2}(\bar r)={2M\Omega^2\over c^2}\left({1\over \bar w}
-{1\over w}\right)={Mc^2\over 2 E^2}\bar w\left(1-{\bar w\over w}\right),
\label{eq:A15}
\end{equation}
where the second equality follows from (\ref{eq:5.7}).

Taking into account  all the orders of the Taylor expansion, one can see
easily from the previous expressions, that the contribution from any
order $\beta$ ($\beta \geq 1$) will be a sum of terms of the form
$ \bar w^\beta \big({w/\bar w}\big)^\alpha$ with $\alpha \le \beta$. When
one substitutes the Taylor expansion (\ref{eq:A13}) of $Q(r)$ into the
integral in (\ref{eq:A8}), such a term will yield the contribution of the form
\[
\bar w\int_{\bar w}^w  \bar w^\beta \left({w\over \bar w}\right)^\alpha
{d w\over w^2}
=\bar w^{1+\beta-\alpha}{w^{\alpha -1}\over \alpha-1}-
{w^\beta\over \alpha-1},
\]
except the case $\alpha=1$, in which case the contribution will have the form
$\bar w^\beta\ln\big\{w/ \bar w\big\}$.
It can thus be seen that the expansion for (\ref{eq:A8}) will take the form
\begin{equation}
  \overline Q(\bar r)=Q(\bar r)-{M\over 4c^2}\Big[{1\over\mu}
{d Q\over d\mu}\Big]_{\bar r}
\bar w\,\ln \big\{ {w_{\!_\odot} \over w} \big\} +{\cal O}\{\bar w\}
\ .\label{eq:A16}
\end{equation}
whose dominant term is not afected be the rigid rotation whose contribution
is proportional to $\Omega$ and thus at most ${\cal O}\{\bar w\}$.

Finally, using once more the Taylor expansion (\ref{eq:A13}), one arrives at
the
simpler form
\begin{equation}
  \overline Q(\bar r)=Q_{_\infty}-{M\over 4c^2}\Big[{1\over\mu}{d Q\over
d\mu}\Big]_{_\infty} \bar w\,\ln \big\{ {w_{\!_\odot}\over \bar w} \big\}
+{\cal O}\{\bar w\} \ .\label{eq:A17}
\end{equation}
We finally obtain a limiting relation of the form
\begin{equation}
\delta_{_\infty}\overline Q\sim \widetilde Q {Mw\over 4c^2\mu^2}\,\ln
    \big\{ {w_{\!_\odot}\over w} \} \ ,\label{eq:A18}
\end{equation}
where the ``gross" asymptotic deviation coefficient is defined
(so as to have the same dimensionality as $Q$ itself) as the logarithmic
derivative
\begin{equation}
\widetilde Q=-\left[\mu {dQ\over d\mu}\right]_{_\infty} \ .\label{eq:A19}
\end{equation}

We now come to the final step, which is to take account of the difference
between the asymptotic limit value $Q_{_\infty}$ with respect to which the
preceeding ``gross" deviations are defined, and the corresponding reference
state value $Q_{_{\ominus}}$ with respect to which the desired ``net"
deviations are defined. In order to proceed, let us use the abbreviation $q$
to indicates some chosen set (which might consist of the relevant
independent longitudinal current components) of the independent field
variables needed to specify a homogeneous state of the system,  so that
(with implicit sumation over the individual variables in the set $q$) the
difference (\ref{eq:A11}) between the actual asymptotic value and its value in
the as
yet unspecified {\it homogeneous reference state} will have an asymptotic
form  given by an expression of the form
\begin{equation}
\delta_{_\infty} Q_{_{\ominus}}\sim{\partial Q\over\partial q}
\,\delta_{_\infty} q_{_{\ominus}} \sim {\partial Q\over\partial q}\,
\big(\delta_{_\infty}\overline{q}-\delta_{_\ominus}\overline{q}\big) \ .
\label{eq:A20}
\end{equation}
More specificly, let us suppose that the set of variables $q$ is chosen so
as to fix a corresponding the choice of the reference state in an obviously
natural way  by simply matching the corresponding average (and therefore
also the total) values, so that  the homogeneous reference state values
$q_{_{\ominus}}$ agree with average values $\overline {q}$ for the vortex
state under consideration, which is expressible in terms of the notation
of (\ref{eq:A10}) just by
\begin{equation}
\delta_{_\ominus}\overline{q} =0\ . \label{eq:A21}
\end{equation}
Subject to this choice, it evidently follows from (\ref{eq:A20}) that
the ``net" variation (\ref{eq:A10}) in which we are ultimately
interested will be given for a generic quantity $Q$ by
\begin{equation}
\delta_{_\ominus}\overline Q\sim\delta_{_\infty}\overline Q
-{\partial Q\over\partial q}
\,\delta_{_\infty}\overline{q} \ .\label{eq:A22}
\end{equation}
It now follows from the application of (\ref{eq:A18}) to the $q$ variables that
there will be a modified {\it net asymptotic deviation coefficient} given by
\begin{equation}
\widehat Q=\widetilde Q-\widetilde{q}\,{\partial Q\over\partial q}
\ ,\label{eq:A23}
\end{equation}
in terms of which the ``net" deviation (\ref{eq:A24}) will be given by the
asymptotic formula
\begin{equation}
\delta_{_\ominus}\overline Q\sim \widehat Q {Mw\over 4c^2\mu^2}\,\ln
\big\{ {w_{\!_\odot}\over w} \} \ .\label{eq:A24}
\end{equation}

\acknowledgments
We wish to thank D. Priou for  carefully reading the manuscript and
his subsequent  comments.

\end{document}